\documentclass[manuscript]{aastex}

\def\bd{\begin{displaymath}}
\def\be{\begin{equation}}
\def\ed{\end{displaymath}}
\def\ee{\end{equation}}

\begin{document}

\title{\bf Analysis of the Positron Fraction and the Spectrum of the Electronic Component in Cosmic Rays}
\author{R. Cowsik and B. Burch\\
\emph{Physics Department and McDonnell Center for the Space Sciences\\
Washington University, St. Louis, MO 63130}}

\begin{abstract}
Recently, the PAMELA, FERMI, HESS, and ATIC instruments have discovered interesting spectral features in the positron to total electron ratio and in the total electronic component of cosmic rays at high energy. These observations are studied with analytical models for the propagation of cosmic rays emanating  from a discrete set of sources or from a smooth distribution of sources extending over the Galaxy. The contrast between the theoretically expected spectra from these two source distributions is seen to play a crucial role in the interpretation of the recent findings. It is shown that the positron to electron ratio, observed by PAMELA, may be fit over the entire energy range in the nested leaky-box model for cosmic ray propagation and that the ratio is expected to reach an asymptotic value of $\sim0.6$ at high energies in all conventional models of cosmic ray propagation. We also derive the spectral shape of the electrons and positrons expected from the annihilation of
 dark matter in the Galaxy and show that the spectral shape of the peak will provide important information, not only regarding the mass of the dark matter particles, but also regarding their spatial distribution. The spectrum of secondary positrons and electrons, calculated with the nested leaky-box model, is subtracted from the spectrum of the electronic component to determine the spectrum of primary electrons emerging from the cosmic ray sources. This spectrum is analyzed in terms of contributions from a set of discrete sources sprinkled across the Galactic volume. Our analysis suggests the possible presence of electrons, accelerated in high Mach number shocks, in the cosmic rays.
\end{abstract}

\maketitle

\section{Introduction}
Direct observation of the cosmic ray electronic component (which includes both electrons and positrons) without any charge discrimination dates back to the early 1960s, and since that time, the energy range and the sensitivity of the observations have increased systematically. To date, we have at hand data from three new instruments, FERMI \cite{FERMI}, ATIC \cite{ATIC}, and HESS (Aharonian et al. 2008, 2009) , that have the requisite sensitivity to measure, with good statistical accuracy, the spectrum of the total electronic component ($e^-+e^+$) well into the TeV region. The reported spectrum in the region $E\geq10$ GeV may be parameterized as
\be\label{eq:spect} F_{t}(E)=I_{t} E^{-\Gamma_t} \ee
with $\Gamma_{t}=3.05$ up to $\sim1$ TeV, and  the value of $\Gamma_{t}$ increases to $\sim$ 3.9 at higher energies.

Furthermore, the ATIC collabortion has reported a narrow enhancement of the intensities around $600$ GeV, superimposed on the smooth spectrum.  We reproduce in fig. 1 their observations and a compilation of the results of other measurements. The spectral slope below $10$ GeV progressively flattens to a slope of $\sim$1.7. Uncertainties in the estimates of the interstellar fluxes are introduced due to solar modulation effects below an energy of a few GeV. We also show, in the same figure, a smooth fit to all the data that we adopt for some of the calculations.

The electronic component in cosmic rays, because of its interactions with radiation fields such as starlight, the microwave background, and magnetic fields in the Galaxy, has been particularly useful in understanding the origins and propagation of energetic particles in the Galaxy \cite{Cowsik66}. This and other early considerations of the effects on the spectral shape of the cosmic ray electrons were carried out in the context of a smooth distribution of cosmic ray sources in the Galaxy, and the transport was described within the framework of the leaky-box model \cite{Cowsik67}. In its initial formulation, the cosmic ray residence time in the Galaxy was assumed to be independent of energy for $E\gg1$ GeV.  However, improved measurements showed that the ratio of secondary nuclei to their parent primary nuclei decreased with increasing energy beyond $\sim 1$ GeV.  This observation naturally gave rise to the suggestion that the residence time of cosmic rays in the Galaxy could indeed be energy dependent and decrease with increasing energy.  According to this model, the anisotropy of the cosmic ray fluxes will correspondingly increase with energy and will exceed the observational bounds at very high energies.  Motivated by the need to alleviate this difficulty and by general considerations, the nested leaky-box model was proposed, which took into consideration the effects of storage of cosmic rays in a small bubble or cocoon surrounding the compact sources of cosmic rays (Cowsik \& Wilson 1973, 1975).  Eventhough the residence time of cosmic rays in the cocoon was energy dependent to generate the energy dependent part of the secondary to primary ratio of cosmic ray nuclei, subsequent to their leakage from the cocoon, the residence time of cosmic rays in the Galaxy was assumed to be independant of energy.  Thus, the anisotropy remained sensibly constant up to very high energies, $\sim 10^{5}-10^{6}$ GeV. We provide some details of this model later in this paper.

The discrete nature of the sources of cosmic rays was addressed within the context of a diffusion model, and the implication to the electronic component was worked out \cite{Cowsik79} to show that we need sources within a few hundred parsecs in order to reproduce the observed spectrum of electrons up to $\sim1$ TeV. Here, the sources of cosmic rays were distributed randomly in a thin disk, and the diffusion region was a thick disk. The boundary condition was that the density of cosmic rays vanished at the upper and lower surfaces of the thick disk, signifying free escape. The need for the proximity of cosmic ray sources to explain the electron spectrum at high energies was reiterated by similar analyses by Nishimura and others (Nishimura et al. 1997; Atoyan et al. 1995). In many of these papers, the complexities introduced by the boundary conditions on the two planes bounding the diffusion volume is avoided by assuming that the diffusion volume is of infinite thickness.

In  this paper, we would also like to keep the Green's function simpler than that provided by Cowsik and Lee (1979), but we would like to still have the option of leakage of the particles from the Galaxy. To this end, we include a term signifying simple leakage similar to that in the leaky-box model. The considerable success of this model in reproducing the energy spectra from the cosmic ray sources and the fact that the Green's function integrates to the leaky-box model for a uniform distribution of sources surrounding the observation point justifies this approach. Apart from the transparency with which the results reported are derived here, we also note that many of the results are indeed generic to any reasonable model of cosmic ray propagation.

It may be noted that this model fits the cosmic ray observations such as the ratio of secondary to primary nuclei in the cosmic rays, the positron fraction, the upper bounds on anisotropy, and the total electron spectrum. Whereas we refer our readers to the papers cited here (Cowsik et al. 1967; Cowsik \& Wilson 1973, 1975) for the full details of the model, we provide here all the details needed for the discussion of the recent data on the electronic component. We show that the spectral slope of the electronic component at the highest energies will be dominated by secondaries generated by the nuclear component of cosmic rays through collisions with the interstellar matter. This is because as long as the cosmic ray sources are discrete, the electrons directly accelerated by them would be cut off sharply due to synchrotron radiation and the inverse Compton effect at energies determined by the distance to the nearest source.

The recent observations of positrons in cosmic rays by PAMELA \cite{PAMELA} has created much excitement because of the possible interpretation of these observations as annihilation of dark matter in the Galaxy.  Their observations and the results from earlier measurements are compiled in fig. 2. Even at the outset, it is worth pointing out that the positron fraction observed by PAMELA has a higher statistical significance and a similar energy dependence in comparison with the earlier measurements, but falls markedly lower over the entire energy band of the measurements. It appears difficult to ascribe the lower ratios observed by PAMELA to the effects of solar modulation, as they are seen even up to $\sim 20$ GeV.   The positron fraction at $\sim1.64$ GeV was measured by PAMELA to be $\sim0.0673$, which decreases to $\sim0.0483$ at $\sim6.8$ GeV and thereupon increases monotonically, reaching a value of 0.137 at a mean energy of $82.55$ GeV. It is this monotonic increase that is being called anomalous, as it does not conform to the prediction of the leaky-box model of the cosmic ray propagation with energy dependent residence time in the Galaxy \cite{Moskalenko}.  Accordingly, this paper begins with a review of the cosmic ray propagation models.   The theoretical expectation that the positron fraction should level off or gently decrease with energy is specific to the assumption that the ratio of secondary to primary nuclei, like B/C, will continue to fall up to very high energies.  In interpreting the positron fraction, it is important to note that there is a fundamental difference between kinematics of the generation of positrons and secondary nuclei by the interaction of primary cosmic ray nuclei; whereas the secondary nuclei emerge from the collisions with nearly the same energy per nucleon as the primary nuclei that generate them, the positron produced through the interactions of cosmic ray nuclei carries away only a few percent of the energy per nucleon of the primary.

This fact leads to the very interesting consequence that the secondary positron spectrum in the interstellar space has the same spectral slope as the primary nuclei in the context of the nested leaky-box model even though the ratio of secondary to primary nuclei, such as B/C, follows the observed decrease with energy.  Of central relevence to the discussion of the positron fraction is the spectrum of the cosmic ray nucleonic component that is dominated by protons with some neutrons coming in bound as He and other nuclei. The nucleon spectrum may be represented as
\be F_n(E)=I_n E^{-\Gamma_n}\ee
where $\Gamma_{n}\approx2.6-2.7$ (We adopt $\Gamma_{n}=2.65$.) in the energy region $1$ GeV/nucleon to $\sim10^6$ GeV/nucleon, beyond which the slope may increase to $\sim$3. The $p/n$ ratio effectively determines the $e^+/e^-$ ratio generated by cosmic rays via interactions with matter in the sources and the interstellar medium through which they propagate before they leak out of the Galaxy. The theoretical calculation of $e^+/e^-$ generated through nuclear interactions of cosmic ray nuclei yield $\sim1.5-2$ \cite{Protheroe}. On the other hand, the observations of the $\mu^+/\mu^-$ ratio \cite{Hayakawa} in cosmic rays gives $\mu^+/\mu^-\approx1.3$.

In recent efforts to explain the positron fraction and the ``bump'' in the total electron spectrum observed by ATIC around $600$ GeV, it has been suggested that astrophysical objects such as a nearby gamma ray burst source \cite{Ioka}, one or more pulsars (Hooper et al. 2008a; Kawanaka et al. 2009; Malyshev et al. 2009; Pohl 2008; Profumo 2008; Y\"{u}ksel et al. 2008), or a few nearby supernova remnants (Dado \& Dar 2009; Fujita et al. 2009; Shaviv et al. 2009) may be the cause of the ``anomaly''. Most of the recent effort has been in looking for a dark matter explanation for these observations. Supersymmetric dark matter such as the neutralino may explain the excess in the observed total electron spectrum, but there may be a need for excessive clumping or an exceedingly local source (Hooper et al. 2008b; Ishiwata et al. 2009). This clumping can be avoided if one considers Kaluza-Klein dark matter \cite{HooperC}. An interesting dark matter explanation comes from the introduction of new force carriers \cite{Arkani} which allows for a Somerfeld enhancement and boosts the annihilation rate to leptons. The new particles may even be able to explain the annual modulation of the signal seen it the DAMA experiment \cite{Bernabei}. Various other dark matter explanations have been formulated to explain the recent findings, and we refer the reader to \cite{Profumo} and \cite{Mertsch} for more detailed lists of the recent studies.

In section 2, we provide a brief overview of the models for cosmic ray propagation and in section 3, we discuss the PAMELA results, first in a model independent way and then compare it with the expectations of the various models. In section 4, we subtract the secondary positron and electron spectra from the spectrum of the total electronic component observed by FERMI, HESS, ATIC, and other experiments to obtain the spectrum of electrons arising exclusively due to input from various discrete sources of cosmic rays sprinkled over the Galaxy. We then analyze this spectrum to derive the distance to the nearest source and the mean spacing between the cosmic ray sources in the Galaxy. In section 5, we analyze the excess in the total electron spectrum by considering a $\delta$-function input spectrum and the spectrum expected from shock acceleration. Finally, section 6 is devoted to a discussion of the main results of this paper and related matters.

This paper is an extension of work presented at the 31st International Cosmic Ray Conferece \cite{Cowsik09}.

\section{Brief Overview of Models of Cosmic Ray Propagation}
It is generally accepted that cosmic rays are accelerated in a large number of discrete sources distributed in the Galaxy, and the cosmic rays propagate from these sources moving along randomly oriented trajectories, akin to diffusion, until they leak away from the Galaxy. During such a propagation, the cosmic rays might interact with matter surrounding the sources, interstellar matter, radiation, and magnetic fields. Any secondaries generated through such interactions, if charged, will be confined by the interstellar magnetic fields and will therefore follow the same kind of random paths as the primaries before escaping from the Galaxy. The various propagation models are characterized by the specific form chosen for the ``vacuum path length distribution'' \cite{Cowsik67} which describes the probability $P(t)$ that the cosmic rays spend in any given region, such as a cocoon surrounding the sources, or in the general interstellar medium before escaping into the intergalactic
 space. The term ``vacuum'' emphasizes the fact that in specifying $P(t)$, one considers hypothetical particles which do not suffer interactions or lose energy during propagation. The effects of these processes are to be computed based on the distribution of paths $P(t)$.

\subsection{The Leaky-Box Model}
In its simplest original form \cite{Cowsik67}, one assumes that $P(t)$ has a broad distribution with significant amplitude near $t=0$, exemplified by a simple exponential function
\be P(t)=e^{-t/\tau} \ee
where $\tau$ is called the escape lifetime of the cosmic rays. In the original version, $\tau$ was assumed to be sensibly independent of energy beyond $\sim1-2$ GeV. Thereafter, since the discovery that the ratio of the fluxes of secondary cosmic rays to those of the primaries was a decreasing function of energy, $\tau$ was considered to decrease with energy to accommodate the observations. We summarize in fig. 3 and fig. 4 the available observations. The crucial issue here is how $\tau$ behaves at energies beyond $10-20$ GeV where the observations have low statistical significance or are non-existent at much higher energies. Most conventional models today (Moskalenko \& Strong 1998; Strong et al. 2007) assume that
\be\label{eq:TauA}Model~A:~\tau_A(E)\sim\tau_0E^{-\Delta},~~~~~~~~~E>2~GeV/n\ee
with $\Delta\approx0.4-0.5$. Such an extrapolation to high energies is shown as a dotted line in fig. 3 and fig. 4.  To understand the spectra of the nuclear component in this model let the rate at which sources inject cosmic ray nuclei, say carbon, into the galactic volume, per unit volume, unit energy interval, and unit time be given by
\be S_c(E) = K_c E^{-\beta_c} \ee

Assuming that the leakage from the Galaxy dominates their loss, the equilibrium spectrum in the interstellar space is given by
\be F_c(E) = S_c(E)\tau_A(E) = K_c \tau_0 E^{-(\beta_c + \Delta)} = I_c E^{-\Gamma_c} \ee
In order to match the observed spectrum of carbon nuclei with $\Gamma_c \approx 2.65$, we should choose $\beta_c=\Gamma_c-\Delta\approx2.15$, i.e. a relatively flat injection spectrum.

Now, secondary nuclei, like B, are generated at the same energy per nucleon as their progenitor C, at a rate
\be S_B(E) = cN_H\sigma_{B,C}(E) F_c(E) \ee
where $N_{H}$ is the density of target nucleon in the interstellar medium.  The B production cross section $\sigma_{B, C} (E)$ is nearly independent of energy and
\be S_B(E) = K_B E^{-\Gamma_c} \ee
where $K_{B}= c N_{H}\sigma_{B,C}(E)K_{c}\tau_{0}=cN_{H}\sigma_{B,C}I_{C}$.  The equilibrium sepctrum of B is given by
\be F_B(E) =S_B(E) \tau_0 E^{-\Delta} = K_B \tau_0 E^{-(\Gamma_c + \Delta)} = c N_H \sigma_{B,C} K_c \tau_0^2 E^{-(\beta_c + 2 \Delta)}\ee
Comparing this expression for $F_{B}(E)$ with that for $F_{C}(E)$ in equation 6, one sees that $F_{B}/F_{C}$ ratio falls as $E^{-\Delta}$.  In other words, the ratio of secondary nuclei to their primaries just follows the energy dependence of the leakage lifetime in the leaky-box model.  It is worth noting that in this model the anisotropies in the cosmic ray fluxes increase as $E^{\Delta}$, becoming unacceptably large for $E\sim 10^{4}$ GeV.

In an attempt to overcome this difficulty we may assume that the residence time follows the energy dependence of the secondary to primary ratio, say up to $\sim 10-20$ GeV and levels off theeafter to a constant value.  That is, we may assume that $\tau(E)$ becomes nearly constant at high energies
\be Model~B:~\left\{\begin{array}{ll}
    \tau_B(E)\sim\tau_A(E), & ~~~~~~~~E_B\sim10~GeV\\
    \tau_B(E)\sim\tau_G, & ~~~~~~~~E\gg E_B
    \end{array}\right.\ee
where $\tau_G$ is a constant. This is shown as a solid line in fig. 3 and fig. 4. The constancy of $\tau\sim\tau_G$ at high energies predicts a lower level of anisotropy of cosmic rays without any increase with energy.  However, in this model the spectral slope of the primary nuclei, $\Gamma_{n}=\beta_{n}+\Delta$, for $E<10$ GeV, and $\Gamma_n=\beta_{n}$ for $E \gg10$ GeV.  Thus in order to match the constant value of $\Gamma_{n}\approx 2.65$, we will have to assume that $\beta_{n}=2.65-\Delta\approx 2.15$ for $E<10$ GeV, and $\beta_{n}=2.65$ for $E\gg10$ GeV.

We will refer to the two models described briefly here as leaky-box model A and leaky-box model B respectively.

\subsection{The Nested Leaky-Box Model}
An alternate way of accommodating the falling secondary to primary ratio is in the context of the nested leaky-box model (Cowsik \& Wilson 1973, 1975). Here, one assumes that subsequent to acceleration, cosmic rays at lower energies spend some time in a cocoon-like region surrounding the sources, interacting with the matter there to generate some of the secondaries. The residence time $\tau_s$ in the source region is assumed to be energy dependent, decreasing with increasing energy. On the other hand, once these cosmic rays enter the general interstellar medium, their subsequent transport becomes independent of energy and the residence time becomes equal to $\tau_G$.
\be Model~C:~\left\{\begin{array}{ll}
    \tau_s(E)\sim\tau_B(E)-\tau_G & ~~~~~~~~1~GeV<E<10~GeV\\
    \tau_s(E)\sim\tau_G & ~~~~~~~~10~GeV<E<10^6~GeV
    \end{array}\right.\ee
The net effect of the interactions in these two regions is to generate the correct ratio of the fluxes of secondary nuclei to those of their primaries, as shown in fig. 3 and fig. 4 marked with labels C.

Let the rate at which a cosmic ray source accelerates primary nuclei and injects them into a cocoon surrounding the source  be represented as $s_{n}(E)$ and given by
\be s_n(E) = q_n E^{-\alpha_n} \ee
The total number of cosmic rays within the cocoon is determined by the balance between the injection rate and the leakage characterized by the lifetime $\tau_s(E)$.
\be f_n(E) = s_n(E)\tau_s(E) = q_n E^{-\alpha_n} \tau_s(E).\ee
The nuclear component leaks out of the cocoon at a rate that is inversely proportional to the residence time $\tau_{s}(E)$.  Thus each source injects cosmic ray nuclei into the interstellar medium at a rate given by
\be f_n(E) / \tau_s(E) = s_n(E) = q_n E^{-\alpha_n} \ee
If the spatial number density of the cosmic ray sources is represented by $\nu$, then the injection rate per unit volume is given by
\be S_n(E) = \nu q_n E^{-\alpha_n}. \ee
Note that this spectral exponent $\alpha_{n}$ is identical to that of the input into the cocoons.  In the above analysis, we have neglected the slight decrease in the value of $q_{n}$ due to nuclear interactions with the mateiral in the cocoon; more exact expressions are available elsewhere (Cowsik and Wilson 1973, 1975), which are also applicable to the rest of the discussions in this section.   Now, the leakage lifetime from the Galaxy $\tau_G$, in the nested leaky-box model, is independent of energy, and the equilibrium spectrum of primary nuclei $F_n(E)$ in the Galaxy is given by
\be F_n(E) = S_n(E)\tau_G = \nu q_n \tau_G E^{-\alpha_n} \equiv I_n E^{-\Gamma_n}, ~~~~~i.e.~~ \alpha_n = \Gamma_n \ee
Note that this represents the spectral form intrinsic to the acceleration process. In constrast with this result for primary nuclei, note that the rate of generation of secondary nuclei within the cocoon, such as B, by the spallation of primary nuclei, such as C, is given by
\be s_B(E) = c n_H \sigma_{B,C} f_c(E) = c n_H \sigma_{B,C} q_c \tau_S(E) E^{-\alpha_c} \ee
This leads to a spectrum of secondary nuclei inside the cocoon $f_{B}(E)$ given by
\be f_B(E) = s_B(E) \tau_S(E) = c n_H \sigma_{B,C} q_c \tau_S^2(E) E^{-\alpha_c} \ee
and the effective rate of injection of secondary nuclei by the cocoons into the insterstellar medium becomes
\be S_{B1}(E) = \nu f_c(E)/\tau_S(E) = c n_H \sigma_{B, C} \nu q_c \tau_S(E) E^{-\alpha_c}. \ee
This rate of injection is just due to the spallation taking place in the sources and reflects the energy dependence of $\tau_{s}(E)$ superimposed on the primary spectrum $\sim E^{-\alpha_{c}}$.  The spallation of primary nuclei in collision with interstellar matter also generates secondary nuclei at the rate
\be S_{B2}(E) = c N_H \sigma_{B,C} F_c(E) = c N_H \sigma_{B,C} \nu q_c \tau_G E^{-\alpha_c} \ee
These two injection processes for secondaries will lead to an equilibrium spectrum of secondary nuclei, $F_{B}(E)$,
\be F_B(E) = \{S_{B1}(E) + S_{B2}(E)\} \tau_G = c \sigma_{B, C} q_c \nu \tau_G \{ n_{H} \tau_s(E) + N_H \tau_G \} E^{-\Gamma_c}. \ee

As noted before, these remarks are made with the assumption that the production of the tertiary nuclei and higher order effects are small. However, it is straightforward to derive the exact expression in a more general case (Cowsik \& Wilson 1973, 1975).  Note that the expression \{$n_{H}{\tau}(E)+ N_{H} \tau_{s}$\} in equation 21 has been chosen so as to reproduce the observed B/C ratio in cosmic rays.

It is important to note that in model C, the spectrum of the primary nuclei as observed by various experiments has the same spectral index as that generated by the acceleration process, both being $\sim-2.65$. The absence of significant anisotropy of cosmic rays at high energies is also easily understood. The residence time of cosmic rays in the Galaxy is essentially constant up to $\sim100$ TeV. This will predict that anisotropy will not increase significantly with energy at least up to this energy.

\subsection{Spectrum of Secondary Electrons and Positrons in Cosmic Ray Propagation Models}
The generation of electrons and positrons in the interactions of the cosmic ray nuclear component occurs through the production of mesons, mainly pions, which decay to muons, which in turn decay into electrons or positrons, transferring, on the average, a fraction of about 0.05 of the energy per nucleon of the primary. This is in contrast with the production of secondary nuclei, such as boron from the collision of carbon nuclei, where boron emerges with almost the same energy per nucleon as the primary carbon nucleus. This difference in their production characteristics leads to nearly identical source spectra $S_{n^-}$ and $S_{n^+}$ for the secondary electrons and positrons $\sim E^{-\Gamma_{n}}$ in all the three models (A, B, and C).  To see this, we note that the rate of generation of secondary positrons and electrons at any energy E is proportional to the flux density of the nucleonic component at $E_n\approx E/0.05 \approx 20 E$ and the collision rate.  Thus in all the models (A,
  B, and C) the rate of production of positrons and electrons in the interstellar medium is given by
\be S_{n\pm} \propto F_n(20E) \sim E^{-\Gamma_n} \ee

In considering the nested leaky-box model C, we should in principle add the generation of secondaries inside the cocoons.  This rate is proportional to
\be S_{n \pm c}(E) \propto f_n(20E) \sim q_n E^{-\alpha_n} \tau_S(20E) \rightarrow 0 ~~~~E\gg1~GeV \ee
In writing equation 23, we have used equation 13 to define $f_n$ and note from fig. 3 and fig. 4 that $\tau_ s$ becomes very small compared with $\tau_{G}$ for $E_{n}=20E\gtrsim 20$ GeV.  Thus the injection spectra of positions and electrons for $E> 1$ GeV are essentially the same in all the three models.

On the other hand, their equilibrium spectra $F_{n^+}(E)$ and $F_{n^-}(E)$ are markedly different in the three models. For energies below $\sim100$ GeV, where the energy losses due to synchrotron radiation and inverse Compton scattering on radiation fields are not important, the three models generate the spectra noted below:
\begin{eqnarray}\label{eq:fn}
Model~A:~~F_{n^+}&\sim& S_{n^+}(E)\tau_A(E)\sim\tau_0E^{-(\Gamma_n+\Delta)}\\
Model~B:~~F_{n^+}&\sim&S_{n^+}(E)\tau_B(E)\sim\tau_0E^{-(\Gamma_n+\Delta)},~~~~~~~E<10~GeV\nonumber\\
&\sim&S_{n^+}(E)\tau_G\sim E^{-\Gamma_n},\quad\quad~~~~~~~~~~~~~~E>10~GeV\nonumber\\
Model~C:~~F_{n^+}&\sim&~S_{n^+}(E)\tau_G\sim E^{-\Gamma_n},\quad ~~~~~~~~~~~~~~~~E>1~GeV\nonumber
\end{eqnarray}

The spectra for the electrons are similar to those given in equation 24, except that because of the dominance of the protons in the cosmic ray beam, the production rate of electrons is somewhat lower, with
\be \frac{S_{n^-}(E)}{S_{n^+}(E)}=\eta.\ee
This ratio $\eta$ is theoretically estimated from the characteristics of high energy interactions to be $\sim$0.5 \cite{Moskalenko}; on the other hand, the direct observation of the $\mu^-/\mu^+$ ratio indicates a value of $\sim0.8$ \cite{Hayakawa}. In either case, $\eta$ is essentially independent of energy beyond a few GeV.

Thus we see that, at high energies ($E\gg10$ GeV), in the leaky-box model B and in the nested leaky-box model C, the secondary positron and electron spectra are power laws with indices $\Gamma_e=\Gamma_n$, i.e. equal to that of the spectrum of the nuclear component in cosmic rays. At very high energies, the energy losses due to synchrotron emission and inverse Compton scattering will steepen these spectra to $F_{n^+}=E^{-(\Gamma_{e}+1)}$. We will discuss this aspect further in section 4.

\section{Analysis of the Positron Fraction Observed by PAMELA}
It is useful to write the observed positron fraction $R(E)$ in terms of the various components:
\be R(E)=\frac{F_{n^+}}{F_{n^+}+F_{n^-}+F_{e^-}}.\ee
Here, $F_{n^+}$ and $F_{n^-}$ represent the positron and electron spectra generated as secondaries of the nuclear component of cosmic rays, and $F_{e^-}$ is the spectrum of electrons resulting from direct acceleration in the sources. Note that there is no direct acceleration of positrons in the source. It is convenient sometimes to work with the inverse of $R(E)$ given by
\be P(E)=\frac{1}{R(E)}=\frac{F_{n^+}+F_{n^-}+F_{e^-}}{F_{n^+}}= 1+\eta+\frac{F_{e^-}}{F_{n^+}}.\ee
This allows one to find the spectrum of electrons generated by the sources $F_{e^-}$ as
\be F_{e^-}(E)=[P(E)-(1+\eta)]F_{n^+}(E).\ee
We show in fig. 5 the net secondary spectrum $F_{n^\pm}(E)=(1+n)F_{n^+}(E)$ calculated in model C, along with the spectrum of electrons resulting from direct acceleration in the sources, $F_{e^-}(E)$, which is obtained by simply substracting $F_{n^\pm}(E)$ from each of the observed data points displayed in fig. 1.  We also show in the same figure a solid line representing $F_{e^-}(E)$ calculated using equation 28 and a smooth fit to the PAMELA observations; this line and the data points in fig. 5 represent the spectrum of electrons generated exclusively through acceleration in various cosmic ray sources in the Galaxy.

Alternatively, we may just assume the functional form for $F_{n^+}(E)$ given by the various propagation models and calculate the positron fraction by dividing $F_{n^+}(E)$ by $F_t(E)$, the total spectrum of electrons measured by FERMI, HESS, and other experiments,
\be R_M(E)=\frac{F_{n^+}(E)}{F_t(E)},\ee
which is shown in fig. \ref{PAMMODELS} along with the data from PAMELA and other experiments. The normalization of the theoretical curves is such as to provide the best possible fit to the three models A, B, and C described earlier. (This normalization may indeed be explicitly calculated as it is proportional to $\tau_A(E)$, $\tau_B(E)$, and $\tau_G$ respectively for the three models and depends on the density of matter in the propagation region, the spectral flux of the nuclear component, the cross section for meson production, decay kinematics, etc.) In depicting the three model curves, we have included the effect of the radiative energy losses at high energies.

Comparison of the theoretical expectations of the positron fraction with the PAMELA data indicates that model A provides a rather poor fit to the observations, as already noted by several authors \cite{PAMELA}. A careful calculation of the positron fraction under the general assumptions of model A was carried out over a decade ago by Moskalenko and Strong (1998). Our estimates here are essentially the same as that derived by them. Even though both model B and model C predict nearly identical injection spectra, the equilibrium spectra at low energies in these two models differ drastically with each other. On the other hand, both these models predict identical equilibrium spectra at high energies for $E\gg10$ GeV. The good fit to observed the positron fraction shows that the residence time of cosmic rays is essentially independent of energy for $E>10$ GeV, a constancy that is expected to continue up to $\sim10^5$ GeV.

In choosing between model B and model C, the latter is preferred from considerations of the spectra of primary nuclei as well. This is because for a simple power law input from the sources having a form $S_n(E)\sim E^{-\beta}$, model B would be expected to yield a spectral form $F_n(E)\sim E^{-(\beta+\Delta)}$ below $\sim10$ GeV and $E^{-\beta}$ at higher energies. On the other hand, the observed spectra of all the primary nucleonic components are simple power laws of slope $\sim E^{-2.65}$ with no changes of slope in the tens of GeV region. Thus we conclude that the nested leaky-box model provides good fit with the PAMELA data and is preferred also from consideration of other cosmic ray observations.

In order to derive the behavior of the positron fraction at very high energies, beyond the PAMELA domain, we need to study the possible form of the spectrum of primary electrons generated by the cosmic ray sources, $F_{e,i}(E)$. The key point to keep in mind here is that there exists a discrete set of compact cosmic ray sources, which are distributed randomly over the Galaxy. This discrete nature of the sources and its impact on the form of the electron spectrum from a source $F_{e,i}(E)$ have been discussed earlier, notably by Cowsik and Lee (1979) and by Nishimura et al. (1997). The diffuse flux from a transient source at a  distance $r$ peaks at a time $t\approx r^2/6\kappa$, where $\kappa$ is the diffusion constant.  This finite time of propagation induces a sharp cutoff in $F_{e,i}(E)$ at high energies due to radiative losses \cite{Cowsik77}. On the other hand, the source functions $S_{n^+}$ or $S_{n^-}$, which correspond to the product of cosmic ray flux and the density of matter in the interstellar medium, have  smooth spatial distributions, and accordingly, lead to a simple steepening of the spectrum to the form $\sim E^{-(\beta+1)}$. Thus, at the highest energies, the secondary component $F_{n^+}(E)$ and $F_{n^-}(E)$ will dominate over $F_{e,i}(E)$, the spectrum of electrons generated by the sources, so that
\be R(E)\sim\frac{f_{n^+}(E)}{f_{n^+}(E)+f_{n^-}(E)}=\frac{1}{1+\eta}\approx0.6.\ee
Thus we expect, on very general considerations, $R(E)$ to increase at high energies and reach a plateau at a level of $(1+n)^{-1}\sim0.6$, which is dictated by the $p$ to $n$ ratio in the interstellar cosmic ray flux and by the nature of nuclear interactions at very high energies.  Based on the discussions in the section 4, we expect the positron ratio to reach the asymptotic value at $ E\gtrsim 5$ TeV.

\section{The Spectrum of Primary Electrons Generated by Cosmic Ray Sources}
For the purposes of this discussion, we write down the transport equation for cosmic rays as
\be\label{eq:trans} \frac{dN}{dt}-\kappa\nabla^2N+\frac{N}{\tau}=Q \ee
where $N$ is the number density of cosmic ray particles at a distance $r$ from the source at time $t$, $\kappa$ is the diffusion constant, $\tau=\tau_G$ is the mean lifetime for the escape of particles from the Galaxy (taken to be independent of energy in the nested leaky-box model), and $Q$ is the source term. Setting $Q=\delta(t)\delta(r)$, we get the Green's function $G(r,t)$ for the transport
\be\label{eq:transgreens} G(r,t)=(4\pi\kappa t)^{-3/2}exp\bigg\{-\frac{r^2}{4\kappa t}-\frac{t}{\tau}\bigg\}.\ee
Note that in writing equation 31, we have not included the term representing the energy loss for the electrons $\nabla\cdot\big(bE^2N\big)$. This is because the energy loss due to synchrotron and Compton processes takes away the energy of the electrons in small steps so that the energy loss may be treated as continuous without any stocasticity.  This loss is customarily described by the equation
\be\label{eq:eloss}\frac{dE}{dt}=-bE^2\ee
or
\be\label{eq:elosstime} t=\frac{E(t=0)-E(t)}{bE(t=0)E(t)}=\frac{E_0-E}{bE_0E}\ee
where $E_0$ is the energy $E(t=0)$ and $E\equiv E(t)$. Thus, with the subsidiary condition given in equation 33, the effect of energy loss may be completely taken into account (Cowsik \& Lee 1979). We note also that for a uniform distribution of sources surrounding the observation point, this Green's function leads to the leaky-box model, with an exponential ``path length distribution''
\be\label{eq:leaky}P(t)=\int G(r,t)4\pi r^2dr\sim e^{-t/\tau}.\ee

Consider now a source located at a distance $r_1$, which accelerates particles continuously to a spectrum of the form
\be Q_1(E) = Q_e E^{-\beta}\quad for~E<E_x.\ee
The spectrum that will be observed is given by
\be \label{eq:obsspect} F_d(E,r_1)=\int^{\frac{1}{bE}-\frac{1}{bE_x}}_{0} Q_e E^{-\beta}(1-bEt)^{\beta-2} G(r_1,t)dt.\ee
The conceptual meaning of equation 37 is clear: equations 33 and 34 imply that the energy of the electron or positron at the instant of production, $E(t=0)$, is given by
\be E(t=0)=\frac{E}{1-bEt}\ee
and a unit bandwidth in energy at $E(t=0)$ gets compressed to a bandwidth $(1-bEt)^2$, enhancing the flux density per unit energy interval by this factor. The upper limit of $t=(1/bE-1/bE_x)$ coincides with $E(t=0)=E_x$, the maximum energy up to which the sources accelerate cosmic ray electrons. For the present purposes, we may take $E_x\sim\infty$.

Before we proceed further, we should fix the value of $b$ to be used. To this end, we note that for $E\gg100$ GeV, the scattering cross section of the starlight by the electrons becomes small, as described by the Kline-Nishina formula. Thus we need to consider the effects of the scattering only on the microwave background at 2.7 K and synchrotron radiation in the magnetic fields in the thick disk, which is the region of cosmic ray confinement. The value of $b$ is given by the formula
\begin{eqnarray}
b &=& {3.22\times10^{-3}(W_{ph}~in~eV/cm^{-3})+7.9\times10^{-5}H^2_{\mu Gauss}}~GeV^{-1}Myr^{-1}\nonumber\\
&\approx&8.5\times10^{-4}+7.1\times10^{-4}\approx1.56\times10^{-3}GeV^{-1}Myr^{-1}.
\end{eqnarray}
We have taken $W_{ph}\approx0.25eV~cm^{-3}$ and $<H>=3\mu G$.

We start with a discussion of the spectra for a smooth distribution of sources such as that for the secondary positrons and electrons in model C; their spectral shape is given by
\begin{eqnarray}
\label{eq:powerdist}F_s(E) &=& \int_0^{1/bE} K_{\pm} E^{-\Gamma}(1-bEt)^{\Gamma-2}e^{-t/\tau}dt\\
&\sim& K_{\pm} \tau E^{-\Gamma}~~~~~~~\quad for\quad E\ll E_c=\frac{1}{b\tau}\nonumber\\
&\sim& K_{\pm} b^{-1} E^{-(\Gamma+1)}\quad for\quad E\gg E_c\nonumber.
\end{eqnarray}
It is this form of $F_{s}$ that we adopted in the theoretical estimates for the secondary positron spectrum and positron fraction in section 2.

The contribution of secondary electrons and positrons $F_{{n}{\pm}}$ to the observed spectrum of the total electronic component $F_t(E)$ in cosmic rays is simply
\be F_{n^\pm}(E) = F_{n^+}(E)+F_{n^-}(E)=(1+\eta)F_{n^+}(E)\ee
where $F_{n^+}(E)$ has the form $F_s(E)$ given by equation 40 and is normalized to yield the correct positron fraction as described in section 2. In order to obtain the spectrum of electrons generated exclusively by the cosmic ray sources, we need to subtract $F_{n}{\pm}(E)$ from the total observed spectrum
\be F_{e^-}(E) = F_t(E)-F_{n^\pm}(E).\ee
We show in fig. 5 $F_{e^-}(E)$ obtained by substracting $F_{n}{\pm}(E)$ from the data from ATIC, HESS, FERMI, and other experiments.

In order to understand the nature and the distribution of the sources of cosmic ray electrons, we first show in fig. 7 the spectrum $F_d(E,r_1)$ for various values of $r_1$ and a power law input with $\Gamma=3$ calculated using equation 37. We note that these spectra are similar to those calculated by Cowsik and Lee (1979), and they appear to fit the observations for $r_1\sim100$ pc. However, the spectrum $F_{e^-}(E)$ has been generated by the inputs from an ensemble of discrete cosmic ray sources in the Galaxy located at distances $r_i$. Assuming that the sources are randomly distributed with a mean spacing equal to the distance to the nearest source, the average distance to the $i^{th}$ source is
\be <r_i>\approx r_1 \sqrt{i}\ee
where $r_1$ is the distance to the nearest source. The theoretical expectation for the equilibrium spectrum generated by the ensemble of cosmic ray sources for a power law input is given by
\be \label{eq:obsspectsum}F_D(E) = \sum_{i=1}^N F_d(E,r_i) = \sum_{i=1}^N\int_{t=0}^{1/bE} \frac{k_i}{E^{\Gamma_s}}(1-bEt)^{\Gamma_s-2} G(<r_i>,t)dt\ee
where N is the total number of sources which contribute to the local cosmic ray density.

The spectrum $F_D(E)$ has been calculated for various values of $r_1$, and the results are displayed in fig. 8 along with the spectrum $F_{e^-}(E)$ estimated  in equation 42 from the observations of FERMI, HESS, PAMELA and other experiments. We note that for the distance to the nearest source of $\sim100-200$ pc, and for a typical separation of $\sim100-200$ pc between the sources, the theoretical calculations match the observations reasonably well, except that they fall short of the intensities beyond a few hundred GeV, leaving behind a narrow spectral feature centered around $600$ GeV. This result, derived here more carefully and with better data, confirms the earlier analysis of Cowsik and Lee (1979) and Nishimura et al. (1997) that the spectrum of the cosmic ray electronic component at high energies shows that there are cosmic ray sources at distances of the order of $\sim100$pc from the solar system and that the typical spacing between them is expected to be the same.

\section{Narrow Spectral Features in the Primary Electron Spectrum}
The narrow spectral feature may be isolated by subtracting from the spectrum of the primary electrons $F_{e^-}(E)$, the expected spectrum $F_D(E)$ due to power law inputs and diffusion from these sources:
\be n_e(E)=F_{e^{-}}(E)-F_D(E) \ee
This is displayed in fig. 9, and this excess or more specifically the sharp spectral features observed by ATIC, has been ascribed \emph{in toto} or in part to products of dark matter annihilation, as mentioned in the introduction. In this section, we discuss this excess in terms of two input spectra, one a $\delta$-function in energy and the other a flat spectrum $\sim E^{-2}$, such as that expected for acceleration at planar shocks of high Mach number.

\subsection{$\delta$-function Input Spectrum}
The injection spectrum is assumed to be of the form
\be Q(t=0,E_a)=\delta(E(t=0)-E_a),\ee
which could in principle represent the spectrum arising from annihilation of dark matter.
Without any leakage from the Galaxy, such a spectrum will evolve as
\be F_{\delta 1}(E,t)= \frac{E_a^2}{E^2}\delta\bigg(\frac{E}{1-bEt}-E_a\bigg).\ee

Note that the above spectrum integrates over $E$ to unity for all values of $t$. Now, for a continuous injection and an energy independent leakage, the spectrum is given by
\be F_{\delta c}(E,E_a)=\frac{1}{bE^2}exp-\bigg(\frac{E_a-E}{bE_aE\tau_G}\bigg).\ee
We display in fig. 10 the spectrum calculated for $E_a=1200$ GeV for a spatially smooth distribution of sources and compare it with the narrow feature $n_e(E)$ estimated in equation 45.  In fig. 11, we provide further examples of the theoretical spectra for various values of $E_a$. Note that $\frac{dF_{\delta c}}{dE}$ is positive for $E<\frac{1}{2b\tau}\equiv\frac{E_c}{2}$ and negative for $E>\frac{E_c}{2}$. Here, $E_c=\frac{1}{b\tau}\approx600$ GeV. Thus, for a $\delta$-function input at $E_a\gtrsim300$ GeV, there is no peak in the observed spectrum at the energy value $E=E_a$. Therefore, the narrow feature seen at $\sim600$ GeV is unlikely to be generated by an input $\delta$-function source spatially extended and smoothly distributed about the solar system. Incidentally note that if the value of $b$ is about a factor of two smaller than our estimate of $1.56\times10^{-3}$ GeV$^{-1}$Myr$^{-1}$, we can get a broad peak at $\sim 600$ GeV, but then the secondary component will
  extending into the HESS region as shown in fig. 5 (dotted line).

Next we consider a $\delta$-function input from single discrete source located at various distances $r_i$. This leads to an observed spectrum
\be F_{\delta D}(E,r_i)=\frac{E_a^2}{E^2}\Big(\frac{bE_aE}{4\pi\kappa(E_a-E)}\Big)^{3/2} exp-\Big(\frac{br_i^2EE_a}{4\kappa(E_a-E)}+\frac{E_a-E}{bE_aE\tau_G}\Big).\ee

We display in fig 12, several examples of $F_{\delta d}$ for different values of $r_i$.  This spectrum displays a peak at
\be E_{peak}\approx\frac{6\kappa E_a}{6\kappa+br_i^2E_a}.\ee
For small $r_i$, this peak will be sharp near $E_a$, but with increasing $r_i$, this peak shifts to lower energies and becomes broader. As before, we sum over the sources at various distances to find their net contribution $F_{\delta D}(E)$ and show this in fig. 13 for different values of the mean spacing between the sources. Such a composite spectrum is quite broad, with a very slight maximum around $200$ GeV and does not provide a good fit to the feature at $600$ GeV.

\subsection{Shock Acceleration}
Acceleration of particles at planar shocks of high Mach number generate spectra which are power laws of index $\sim2$
\be Q_{shock}(E)\sim Q_0E^{-2}~~~~~~for~E<E_x.\ee
The upper cutoff energy $E_x$ and the precise shape of the cutoff depends on the nature of the shock, radiative energy losses, and other such features. Such an input leads to an equilibrium spectrum
\be F_2(E)=\frac{\tau Q_0}{E^2}\bigg(1-e^{-\frac{E_x-E}{bEE_x\tau_G}}\bigg).\ee
This is shown in fig. 14, and it appears to reproduce $n_e(E)$, the difference between the primary electron spectrum $F_{e-}(E)$ and that expected by a set of discrete cosmic ray sources $F_d(E)$.

\section{Discussion and Related Matters}
The main result that emerges from the present analysis is that the nested leaky-box model provides a good fit to the positron fraction observed by PAMELA. The model is also consistent with other observations of cosmic rays. Until good measurements of the positron fraction were available, there was no easy means of choosing amongst various models. The fact that the nuclear secondaries, such as Li, Be, and B, emerge from nuclear interactions with essentially the same energy per nucleon as their parents, C, N, and O, was the main cause for this degeneracy. However, the fact that the positrons carry, on the average, a fraction of only about 0.05 of the energy per nucleon of their nuclear primaries breaks this degeneracy, allowing a choice to be made. Improvements of the measurements of the spectra of both the secondary nuclei and of positrons will help in fixing, more firmly, the parameters of the nested leaky-box model. Note that this model predicts that the $\bar{p}/p$ ratio \cite{PBAR} should nearly be constant at $E\gg10$ GeV. Also, the gamma rays generated by cosmic rays inside the cocoons through the $\pi^\circ\rightarrow2\gamma$ process will have, on the average, a steep spectrum $E^{-\beta}_{n} \tau_{s}(E)\sim E^{-(\Gamma_n+\Delta)}$ at $E_{\gamma}>1$ GeV. In closing, we note that as the centenary of the discovery of cosmic rays is approaching us, the field of cosmic ray studies is becoming more vibrant and is intimately getting connected with wider aspects of research in astrophysics.

The recent measurements of the electronic component of cosmic rays by the PAMELA, HESS, and FERMI groups has opened up many interesting new avenues to discuss cosmic ray propagation. Some of these were discussed or hinted at in this paper. Further discussions will include an anysis of the $\bar{p}/p$ ratio as seen by PAMELA \cite{PBAR}. 

\textbf{Acknolwedgement}:It is a pleasure to acknowledge the detailed discussions with Professor Martin A. Israel and the hospitality at the Raman Research Institute where part of this manscript was drafted.

\begin{figure}\label{TotalE}
 \includegraphics[width=16cm]{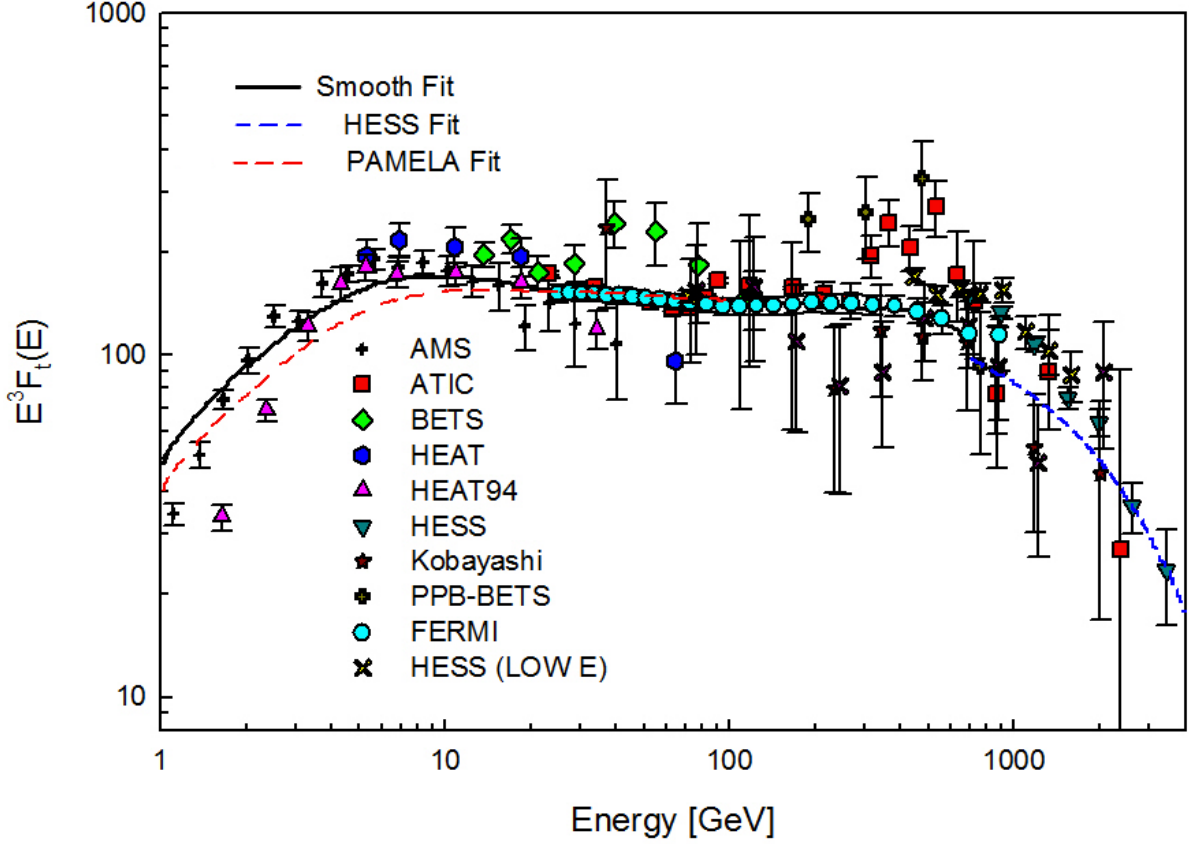}
  \caption{The compilation of measurements of the total electron spectrum $F_t(E)$, the red dashed line represents the total spectrum of the electronic component calculated using the positron ratio measured by PAMELA. The blue dashed line is a fit to the HESS data, and the solid line is a smooth fit to the total electron spectrum.}
\end{figure}

\begin{figure}\label{PAM}
 \includegraphics[width=16cm]{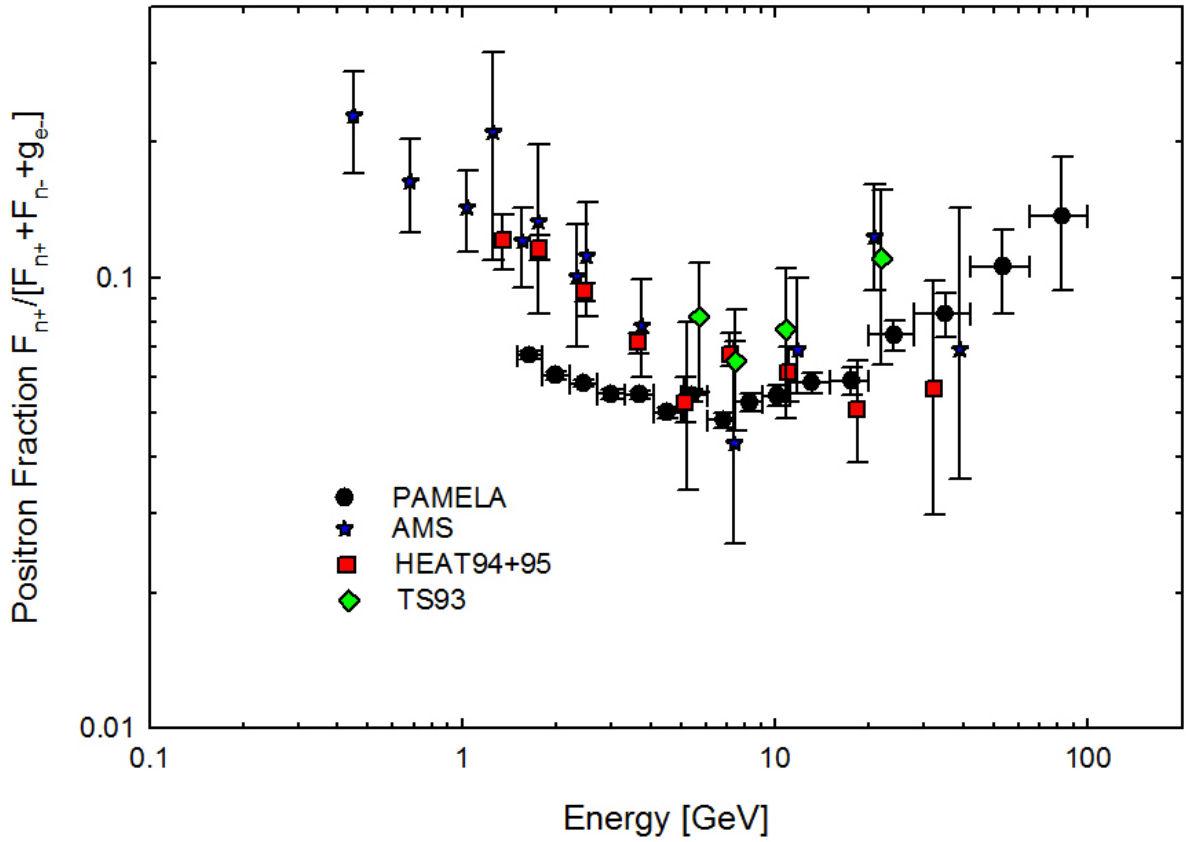}
  \caption{The positron fraction measured by PAMELA along with the earlier measurements are shown. Note that PAMELA data has a similar energy dependence, but is markedly lower than the earlier measurements.}
\end{figure}

\begin{figure}\label{BC}
 \includegraphics[width=16cm]{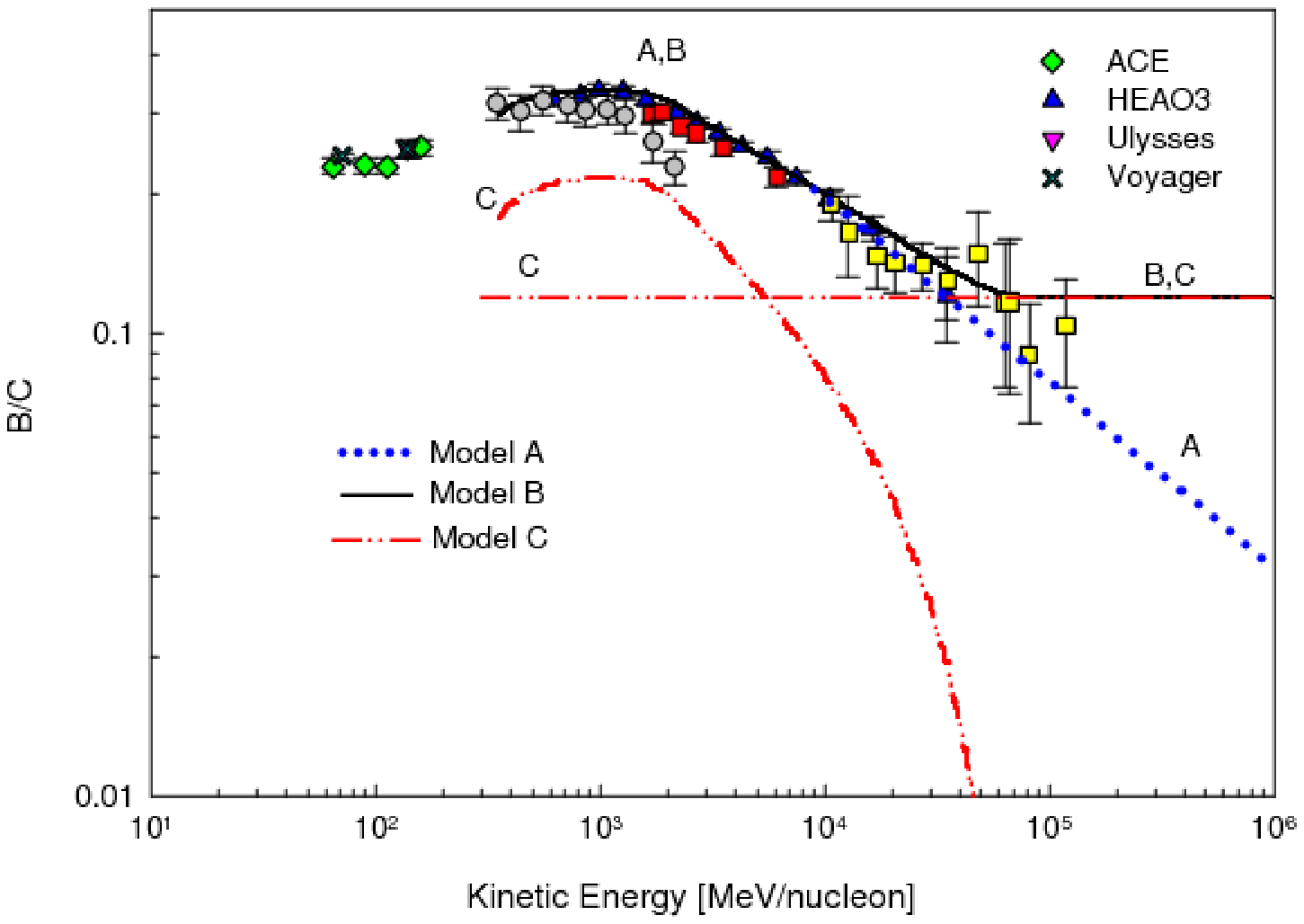}
  \caption{The observed B/C secondary to primary ratio is plotted [points from a compilation in \cite{Strong}] along with the power law extrapolation at high energies (dotted line, Model A), a constant extrapolation (solid line, Model B), and a two-component fit (dot-dashed line followed by solid line for $E\geq50~GeV$, Model C).}
\end{figure}

\begin{figure}\label{SC}
 \includegraphics[width=16cm]{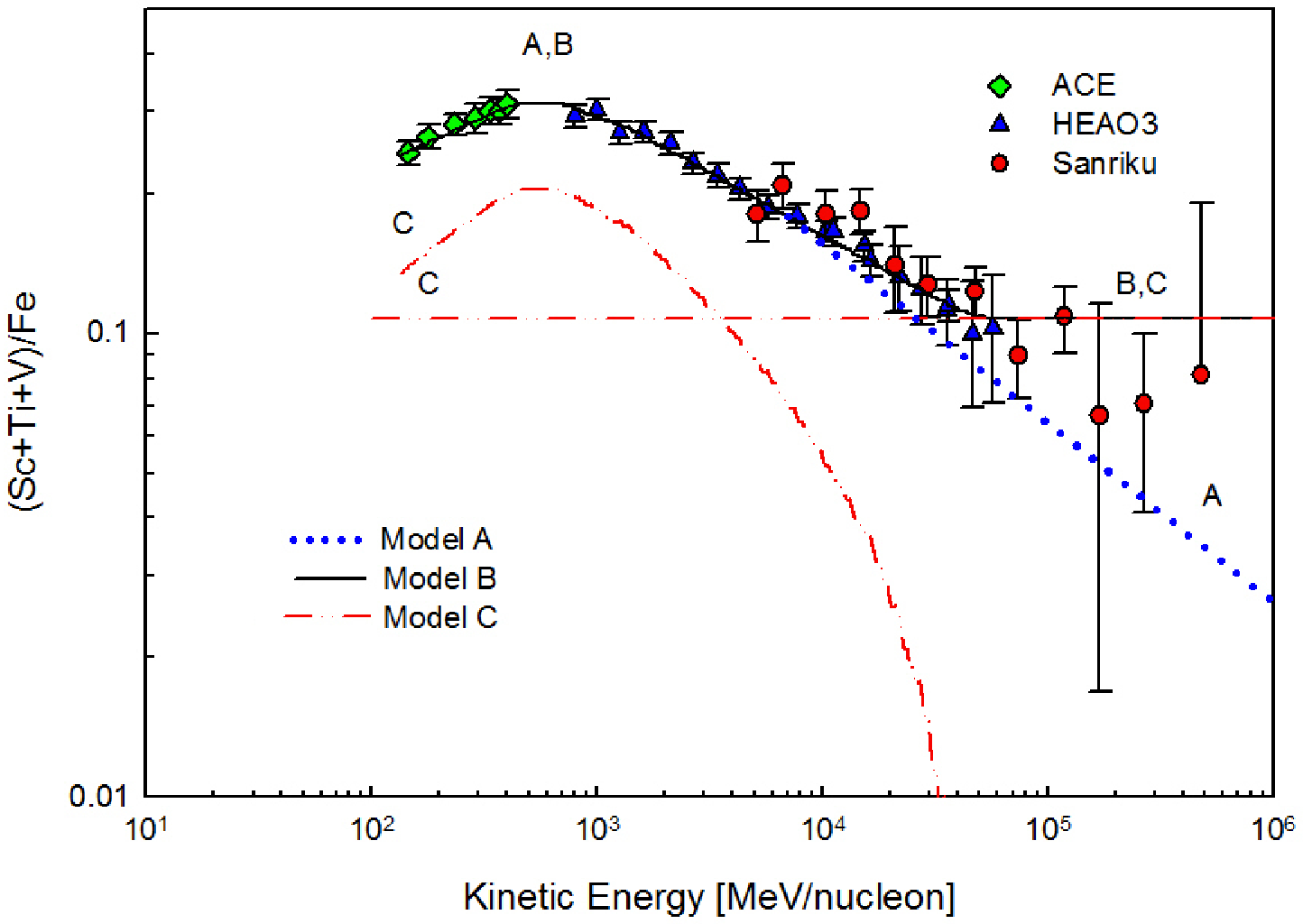}
  \caption{The observed (Sc+Ti+V)/Fe secondary to primary ratio is plotted [points from a compilation in \cite{Strong}] along with the power law extrapolation at high energies (dotted line, Model A), a constant extrapolation (solid line, Model B), and a two-component fit (dot-dashed line followed by solid line for $E\geq50~GeV$, Model C).}
\end{figure}

\begin{figure}\label{PRIMARIES}
 \includegraphics[width=14cm]{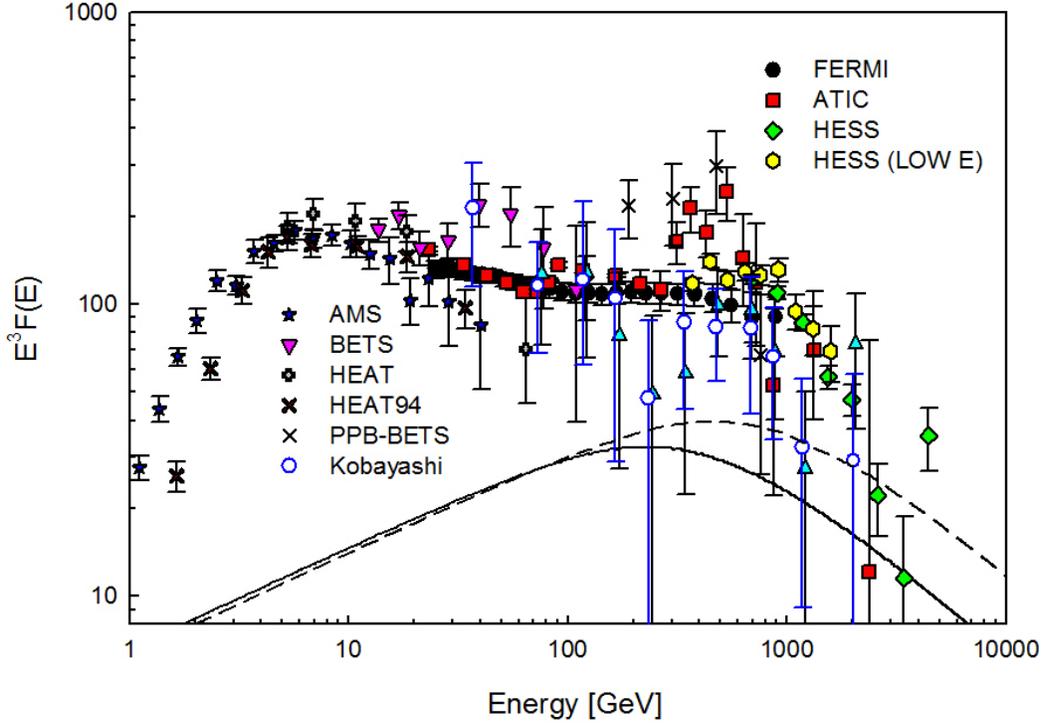}
  \caption{We display here the primary spectrum of cosmic ray electrons $F_{e-}(E)$ generated exclusively through acceleration of electrons in the cosmic ray sources obtained by substracting the secondary positrons and electron (solid line) from the measurements of the total electronic component $F_t(E)$ by HESS, ATIC, FERMI and other experiments.  The sum of the secondary component and $F_n^\pm (E)$ the primary component will add up to $F_t(E)$ showin in fig. 1}
\end{figure}

\begin{figure}\label{PAMMODELS}
 \includegraphics[width=16cm]{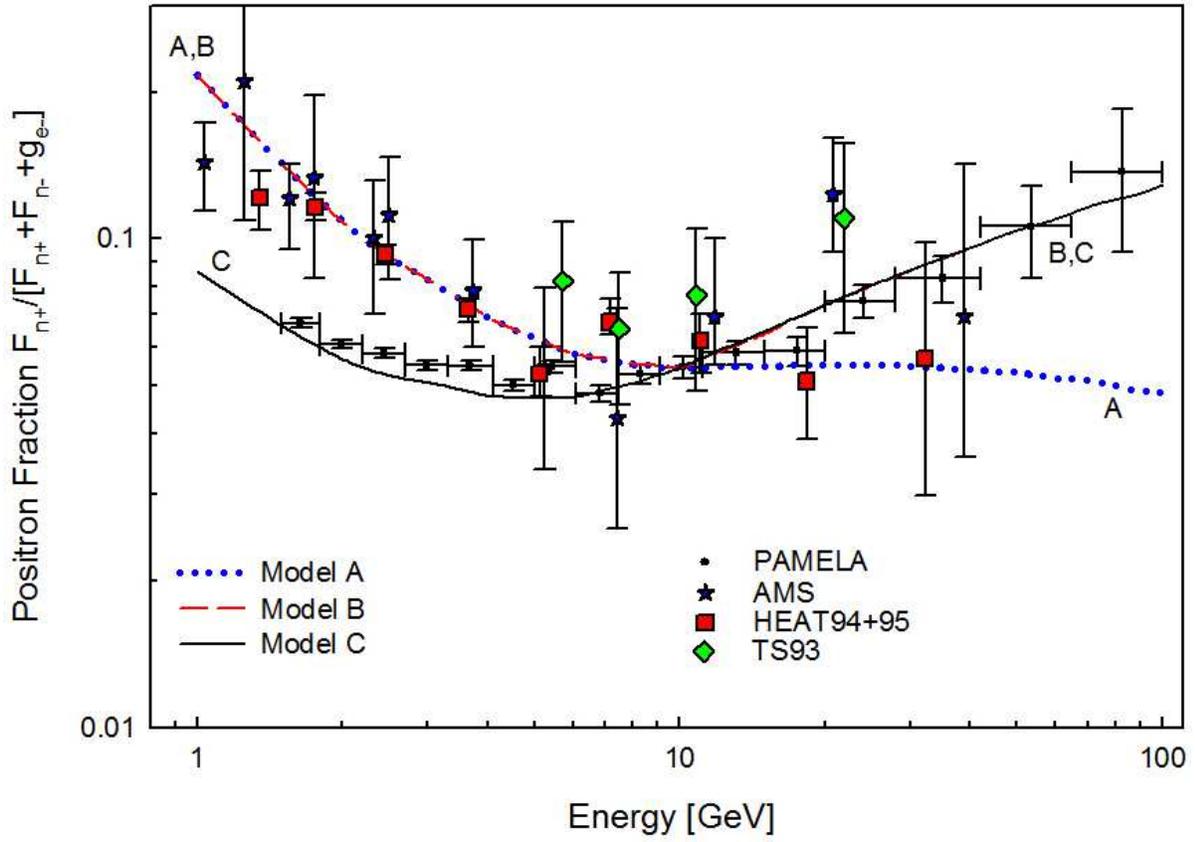}
  \caption{The theoretically calculated positron fraction for models A, B, and C are compared with observations. All calculations are normalized at $\sim10~GeV$, to the PAMELA measurements.}
\end{figure}

\begin{figure}\label{SINGLESOURCE}
 \includegraphics[width=14cm]{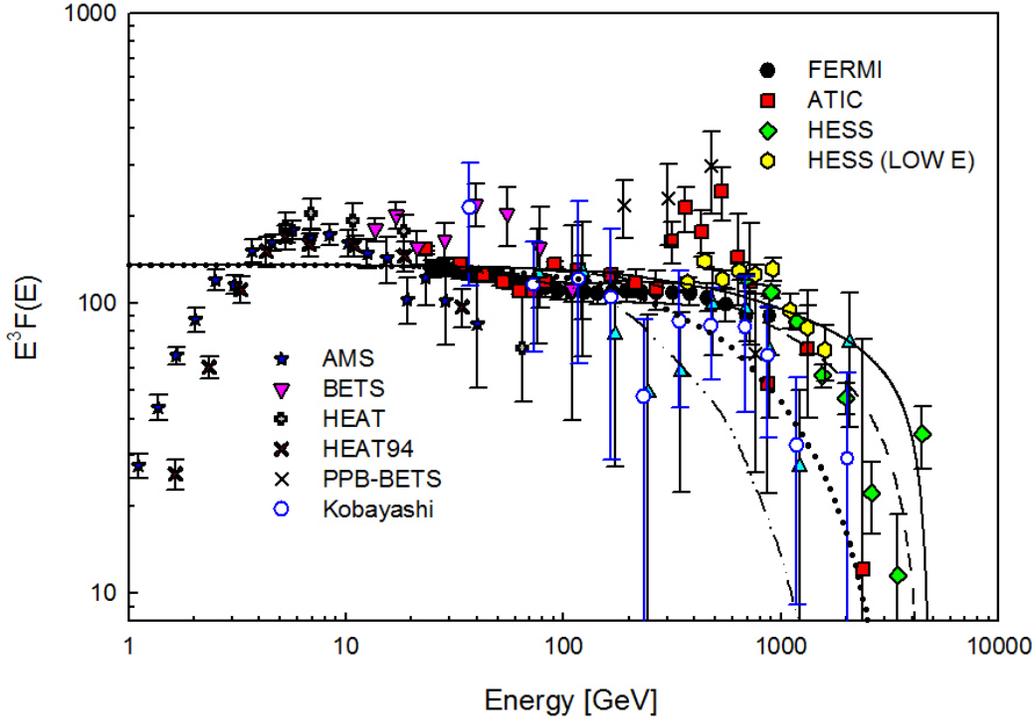}
  \caption{The primary electron spectra due to a single source at various distances from the source with $E_x=5$ TeV compared to the primary electron spectrum. [$r_1=0.1$ kpc (solid line), $r_1=0.2$ kpc (dashed line), $r_1=0.5$ kpc (dotted line), $r_1=1.0$ kpc (dot-dashed line)].}
\end{figure}

\begin{figure}\label{SUMSOURCE}
 \includegraphics[width=14cm]{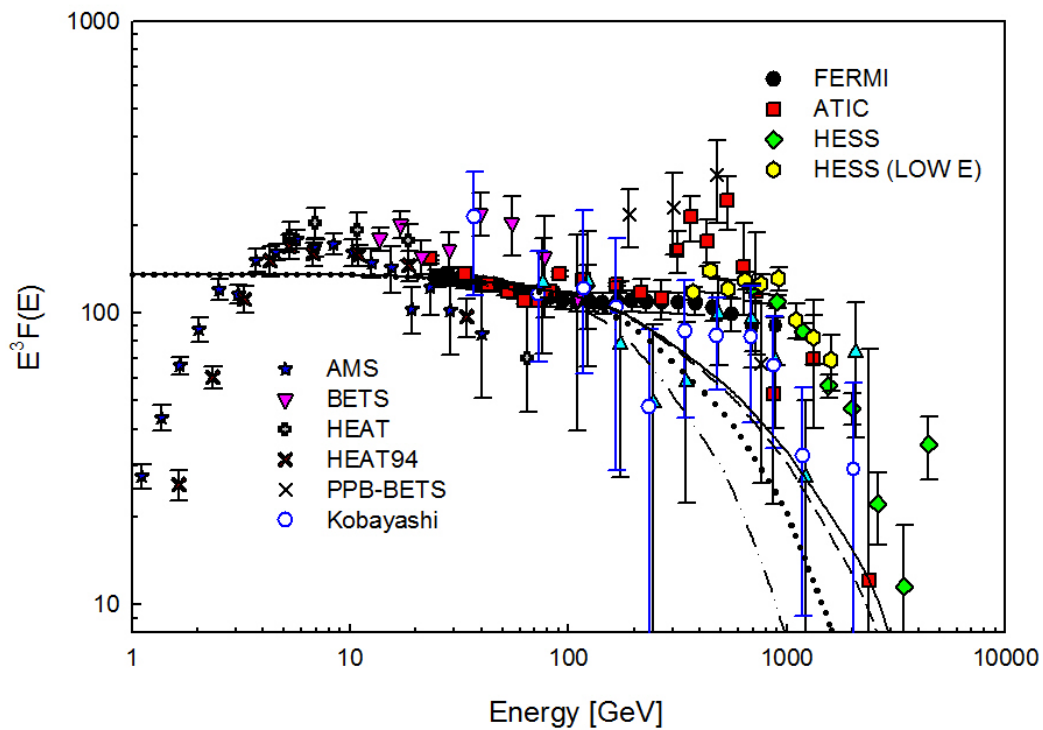}
  \caption{The theoretical primary electron spectra resulting from an ensemble cosmic ray sources for various values of the mean spacing and $E_x =5$ TeV is compared with the primary electron spectrum $F_{e-}(E)$. The mean spacing between the sources is taken to be $<r>=0.1$ kpc (solid line), $<r>=0.2$ kpc (dashed line), $<r>=0.5$ kpc (dotted line), and $<r>_1=1.0$ kpc (dot-dashed line).}
\end{figure}

\begin{figure}\label{LEFTOVER}
 \includegraphics[width=14cm]{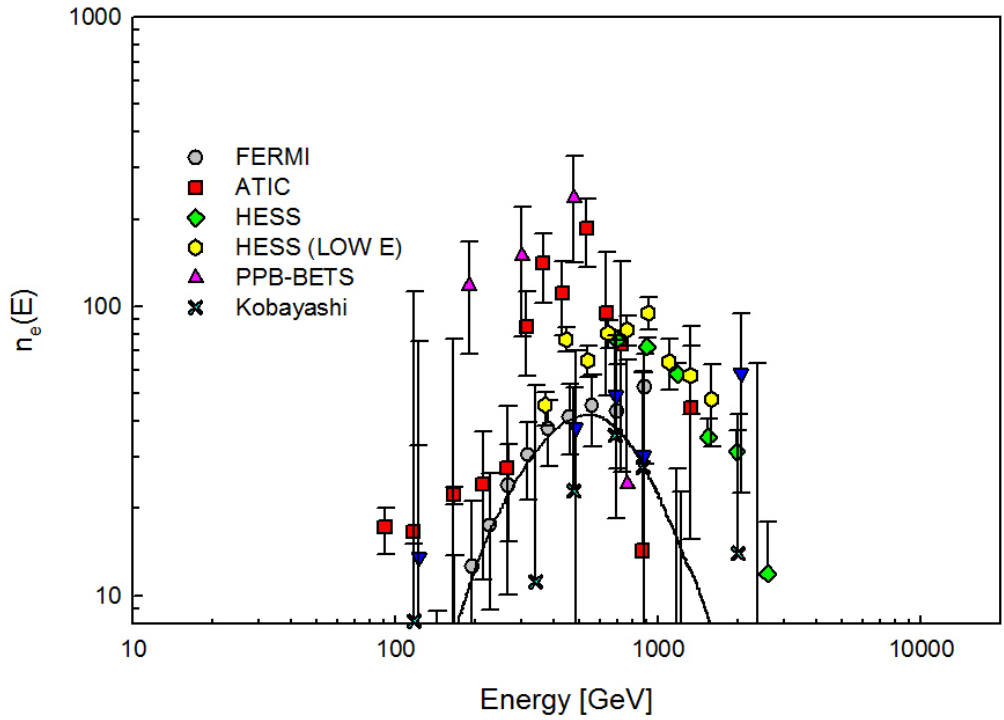}
  \caption{The excess primary electrons obtained after subtracting the expected contribution from discrete sources estimated with a mean spacing of $\sim0.1$ kpc from the primary electron spectrum $F_{e-}(E)$. This is shown both as data points and as  a smooth fit through the data.}
\end{figure}

\begin{figure}\label{DELTACONT}
 \includegraphics[width=14cm]{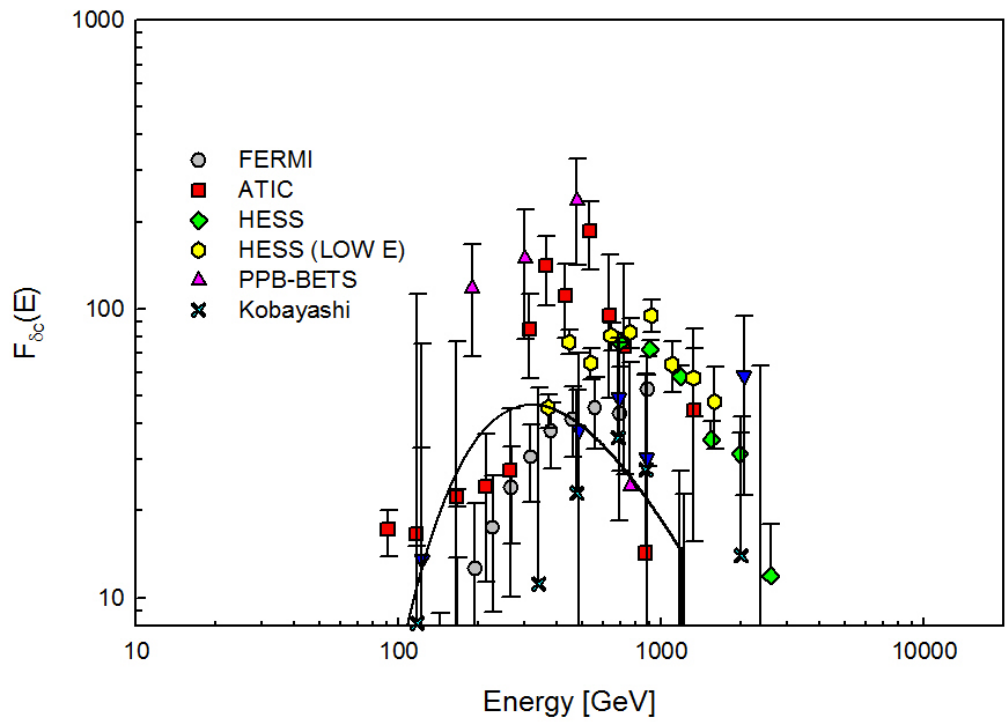}
  \caption{The spectrum for a $\delta$-function input from continuous distribution of sources with $E_a=1200$ GeV . Note the peak at $\sim300$ GeV$\approx E_c/2$ is expected for all $E_a>300$ GeV (see fig. 11}
\end{figure}

\begin{figure}\label{DELTACONTMODELS}
 \includegraphics[width=14cm]{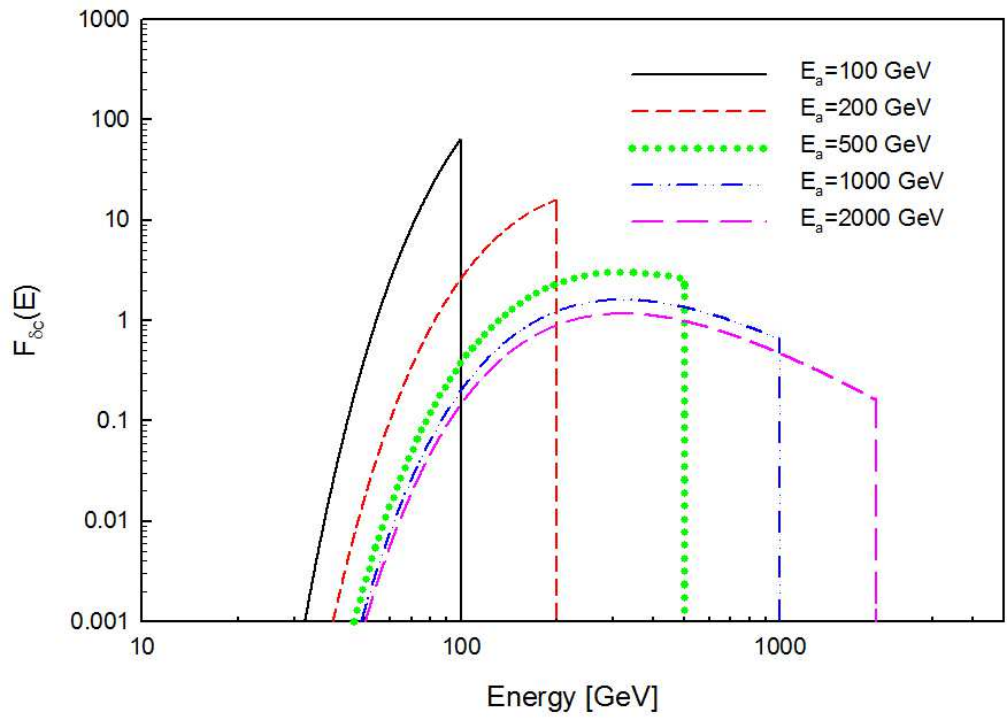}
  \caption{Further examples of the spectra of electrons expected for $\delta$-function inputs from a continuous spatial distribution of sources.}
\end{figure}

\begin{figure}\label{SINGLEDELTA}
 \includegraphics[width=14cm]{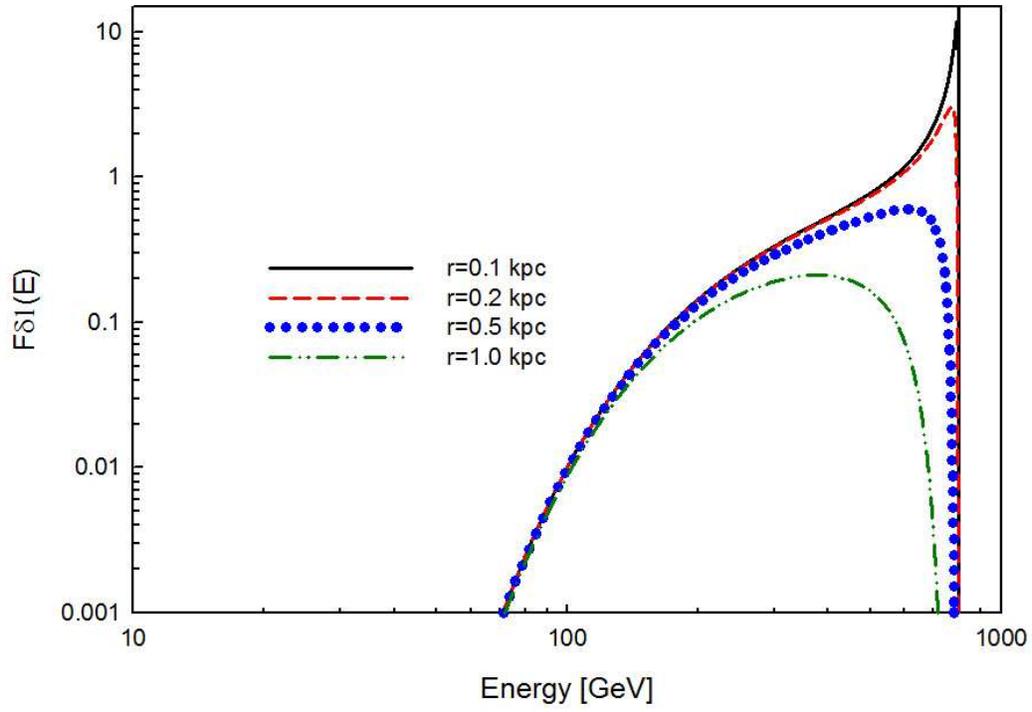}
  \caption{The spectrum of electrons expected for a $\delta$-function input,  diffusing spatially from a single source situated at various distances.}
\end{figure}

\begin{figure}\label{SUMDELTA}
 \includegraphics[width=14cm]{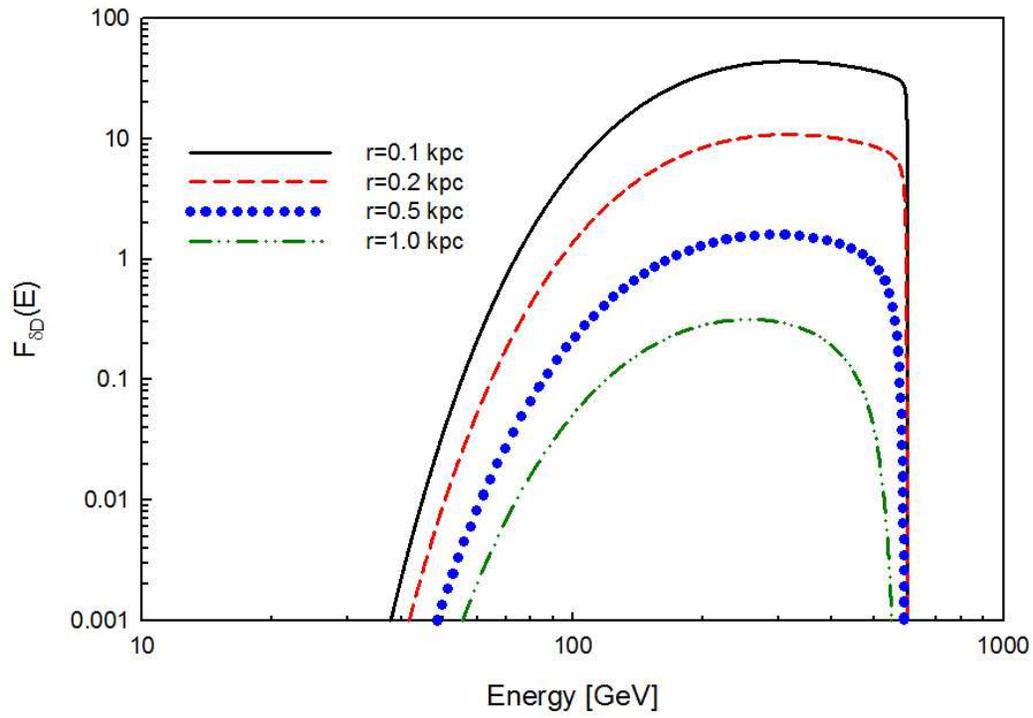}
  \caption{The spectrum of electrons arising due to a $\delta$-function input from a discrete set of sources located at various distances with mean spacing as indicated, and calculated assuming diffusive transport.}
\end{figure}

\begin{figure}\label{SHOCKS}
 \includegraphics[width=14cm]{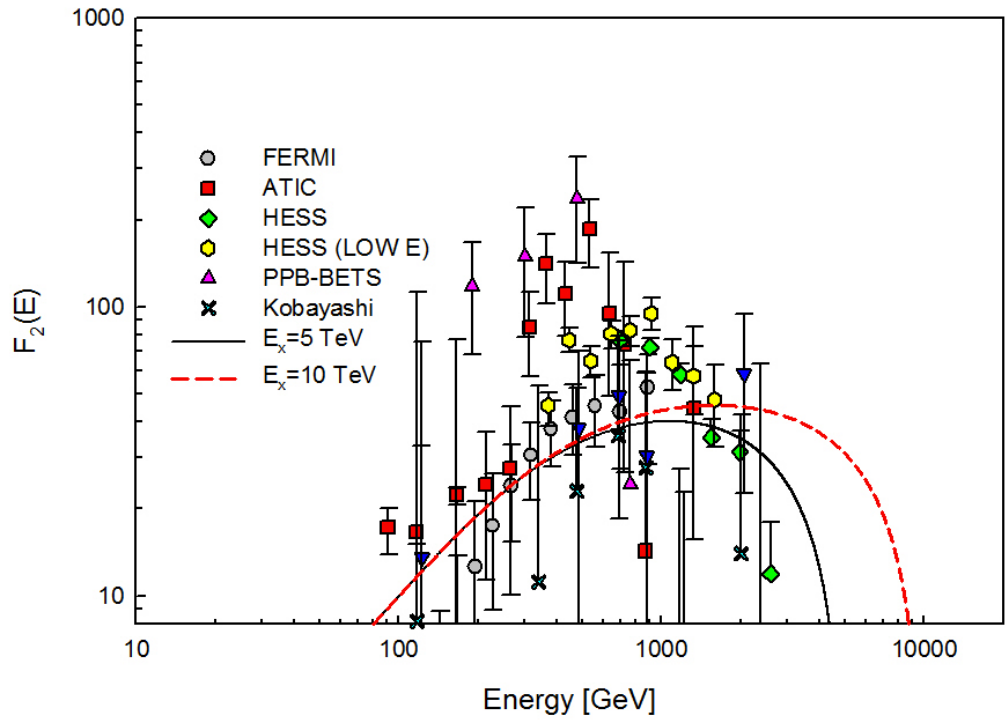}
  \caption{The equilibrium spectrum of electrons arising from shocks to input spectra $\sim E^{-2}$ up to various cutoff energies between $5-10$ TeV is compared with the narrow feature in the observed spectrum.}
\end{figure}

\end{document}